\documentclass{elsart}
\usepackage{epsfig}
\usepackage{epsf}
\begin{document}
\newcommand{\EP}{\mbox{e$^+$}}
\newcommand{\EM}{\mbox{e$^-$}}
\newcommand{\EPEM}{\mbox{e$^+$e$^-$}}
\newcommand{\EMEM}{\mbox{e$^-$e$^-$}}
\newcommand{\EE}{\mbox{ee}}
\newcommand{\GG}{\mbox{$\gamma\gamma$}}
\newcommand{\GE}{\mbox{$\gamma$e}}
\newcommand{\beq}{\begin{equation}}
\newcommand{\eeq}{\end{equation}}
\newcommand{\beqn}{\begin{eqnarray}}
\newcommand{\eeqn}{\end{eqnarray}}
\runauthor{M. Galynskii, E. Kuraev, M. Levchuk, V. Telnov}
\begin{frontmatter}
\title{
Nonlinear effects in Compton scattering at photon colliders}
\author[Minsk]{Michael Galynskii,\thanksref{cor}}
\author[Dubna]{Eduard Kuraev,}
\author[Minsk]{Michael Levchuk,}
\author[Novosibirsk]{Valery Telnov}
\thanks[cor] {Corresponding author, galynski@dragon.bas-net.by}
\address[Minsk]{Institute of Physics BAS, F. Skoryna ave. 68,
220072 Minsk, Belarus}
\address[Dubna]{Joint Institute for Nuclear Research, 141980 Dubna,
Moscow Region, Russia }
\address[Novosibirsk]{Institute of Nuclear Physics, 630090 Novosibirsk, Russia}
\date{}
\begin{abstract}
  The backward Compton scattering is a basic process at future higher
  energy photon colliders. To obtain a high probability of e$\to
  \gamma$ conversion the density of laser photons in the conversion
  region should be so high that simultaneous interaction of one
  electron with several laser photons is possible (nonlinear Compton
  effect). In this paper a detailed consideration of energy spectra,
  helicities of final photons and electrons in nonlinear backward
  Compton scattering of circularly polarized laser photons is given.
  Distributions of $\gamma\gamma$ luminosities with total helicities 0
  and 2 are investigated. Very high intensity of laser wave leads to
  broadening of the energy (luminosity) spectra and shift to lower
  energies (invariant masses). Beside complicated exact formulae,
  approximate formulae for energy spectrum and polarization of
  backscattered photons are given for relatively small nonlinear
  parameter $\xi^2$ (first order correction). All this is necessary
  for optimization of the conversion region at photon colliders and
  study of physics processes where a sharp edge of the luminosity
  spectrum and monochromaticity of collisions are important.

\vspace{0.5cm}
\noindent
PACS: 13.60.Fz; 13.88.+e; 42.60; 12.20.Ds
\end{abstract}

\begin{keyword}
  Laser; Photon photon; Gamma gamma, Electron photon; Backward Compton
  scattering; backscattering; Linear collider; Photon collider
\end{keyword}
\end{frontmatter}
\typeout{SET RUN AUTHOR to \@runauthor}

\section{Introduction}

The process of backward Compton scattering (BCS) is a basic process at
Photon Linear Colliders (PLC), i.e. $\GE$ and $\GG$ colliding beams
based on $\EPEM$ (ee) linear colliders
\cite{KC81,GKST83,NIM84,GKPS83,Akerlof}.  A detailed description of
basic scheme of photon colliders and main characteristics of colliding
$\GE$ and $\GG$ beams can be found in papers \cite{GKST83,NIM84}. In
the latter non-trivial polarization effects in $\GE$ and $\GG$
interactions are also investigated. Further development of these ideas
can be found elsewhere
\cite{TEL90,TEL95,TELvav,TSB2,TSB1,Borden,BERK,NLC,TESLA,JLC,TELee,Tfrei,TGG2000b}.

When a density of laser photons is very high, the processes in the conversion
region can become  nonlinear due to simultaneous
interaction of electrons with several laser photons
\cite{GKP,Ritus,Ritus79,Bula96}:
\beqn
&&e(p)+n\gamma(k_0)  \to  e(p')+\gamma(k)\;,\;\; \quad n \geq 1 \; ,
\label{nelcom} \\
&&\gamma(k)+s\gamma(k_0)  \to  e^{+}(p_+)+e^{-}(p_-)\;,\;s\geq 1\;.
\label{nelpar}
\eeqn 
The first of these nonlinear processes (i.e. a process of
nonlinear BCS) leads to widening energy spectra of scattered high
energy photons and the second one lowers $\EPEM$ pair production
threshold \cite{GKP,Ritus,Ritus79,Bula96}.  An interaction of
electrons and positrons with an electromagnetic wave effectively
increases the mass of the electron and this increase is characterized
by an intensity parameter $\xi^2$:
\beq
m^2 \rightarrow m^2_{\ast} = m^2 \; (1 + \xi^2) \;,\;
\xi^2 = n_{\gamma} \left ({4 \pi \alpha \over m^2 \omega_0} \right )
\; = \; - {e^2 \; a^2 \over m^2} \;  ,
\label{ksi}
\eeq where $n_{\gamma}$ is the photon density in the wave, $\omega_0$
is the energy of the photons, $a$ is the amplitude of the a classical
4-potential of electromagnetic wave, $e$ and $m$ are the charge and
the mass of the electron and $\alpha$ being the fine structure
constant.~\footnote{In this paper we assume $\hbar = 1, c=1$.}  A
systematic theoretical investigation of the nonlinear processes
(\ref{nelcom}) and (\ref{nelpar}) has been done in papers
\cite{Ritus,Ritus79}, and an experimental study has been carried out
in papers \cite{Bula96}.

At photon colliders one has ``to convert'' almost all electrons to
high energy photons. To achieve this with minimum laser energy the
laser light should be optimally focused, namely the length of the laser
bunch and the Rayleigh length (the depth of the laser focus) should be
approximately equal to the electron bunch
length~\cite{GKST83,TEL90,TEL95}.  In this case the transverse size
follows from the transverse emittance of the laser beam which is equal to
$\lambda/4\pi$ (diffraction limit). It turns out that for beam
parameters considered for photon colliders the parameter $\xi^2$ at
optimum focusing may be larger than its acceptable value
$\sim$0.2--0.3. To decrease $\xi^2$ (photon density) keeping
conversion probability constant one has to make the photon bunch at
the focal point longer and wider that requires larger laser flash
energy.  So, nonlinear effects in Compton scattering are very
important for optimization of photon colliders (better quality of
spectra needs larger flash energy).

In this paper we analyze in  detail the influence of nonlinear
effects on main characteristics of $\GE$ and $\GG$ collisions, namely,
the energy spectra of photons, polarization of final photons and
electrons, and the distribution of both the total spectral luminosity
of $\GG$ collisions and that for the total helicity 0 and 2. As we
will see below the nonlinear effects are important. They must be taken
into account at simulations of PLC. For comprehensive simulation of a
PLC \cite{TEL95,TSB2,Borden,Yokoya} including processes multiple BCS
one has to know not only the differential cross section of the reaction
(\ref{nelcom}) but energy, angles and polarization of final photons and
electrons.

\section{The differential cross section for nonlinear BCS}

In the case of head-on collision between ultrarelativistic electrons
and photons of circularly polarized laser waves, the energy dependence
of the differential cross section of process (\ref{nelcom}) as a
function of $y = \omega / \varepsilon$ (where $\varepsilon$ is the
electron energy) has the following form \footnote{The polarization
  states of all particles involved in reaction (\ref{nelcom}) are
  helicity ones.}  \cite{Yokoya,Rek83,GalSik92,Tsai,KGL,GalSik98}:
\beqn {d \sigma_c \over d y}(\lambda, \lambda_e, \lambda', \lambda_e')
& =& {\pi \alpha^2 \over 2 x m^2 \xi^2 } \sum_{n = 1} ^{\infty}
\{\,(1+\lambda_e \lambda_{e}') F_{1n} \nonumber \\
&+&\lambda (\lambda_{e} +\lambda_e')F_{2n} + \lambda'( \lambda
F_{3n}+\lambda_{e} F_{4n})+\lambda_e \lambda_{e}' F_{5n}\,\}\, ,
\label{nely}
\eeqn
\beqn
F_{1n}& =& - 4 \; J_{n}^2 + \xi^2 \; \left ( 1 - y + {1 \over 1 - y } \right ) \;
( J_{n-1}^2 + J_{n+1}^2 - 2 J_{n}^2 ) \; , \nonumber\\
F_{2n}& =& \xi^2 \left ( - 1 + y + { 1 \over 1 - y } \right ) \;
\left ( 1 - 2 {y\over y_{n}}{(1-y_{n}) \over (1-y)} \right )
\; ( J_{n-1}^2 - J_{n+1}^2 ) \; ,  \nonumber  \\
F_{3n}& =& \xi^2 \; \left ( 1-y + {1 \over 1-y } \right ) \; \left ( 1 - 2
\; {y\over y_{n}} {(1-y_{n}) \over (1-y)} \right ) \;
( J_{n-1}^2 - J_{n+1}^2 ) \; ,\label{funy} \\
F_{4n}& =& - 4 y \; J_{n}^2 + \xi^2 \left ( -1+y +{1\over 1-y} \right ) \;
( J_{n-1}^2 + J_{n+1}^2 - 2 J_{n}^2  \; ) \; , \nonumber  \\
F_{5n} &=& 4 J_n^2 \left ( 1+ y -{1 \over 1-y} \right )\;,\;\;
\nonumber\\
z_{n} & =& {2 n \xi \over \sqrt{1+\xi^2}} \; \sqrt{\alpha_{n}}\, ,\,
\alpha_{n} =r_n(1-r_n)= {y \over y_{n}} \left (1 - {y \over y_{n}}\right )
{ (1-y_{n}) \over (1-y)^2}\, , \label{z_n} \\
y_{n}& =& {u_{n} \over 1 + u_{n}}\,, \,r_n={y \over u_n(1-y)} \,,
u_{n}={nx \over 1+\xi^2}\;, \; x = {2 k_0p \over m^2} \;  =
{4 \omega_0 \varepsilon \over m^2} \;,\;
\label{y_n}
\eeqn
where  $J_{n}, J_{n\pm1}$ are the various order Bessel functions of the
same argument $z_{n}$ (\ref{z_n}): $J_n=J_n(z_n)$. Here the $y=\omega/
\varepsilon$ changes in the range $0 \leq y \leq y_{n}$. The expression,
standing in front of the sum (\ref{nely}) gives the probability
for the $n$-th harmonic to be radiated in the case when polarization states
of both initial and final electrons and photons are  $\lambda_e=\pm1,
\lambda_e'=\pm1, \lambda=\pm1, \lambda'=\pm1$.

It should be noted that with backward scattering (when $y=y_n$), all the
functions $F_{in}$ are zeroes at $n>1$. This means that only the first
harmonic photons can be radiated in the direction along that of the
initial electron beam. Higher harmonic photons can not be radiated in such
conditions because of the helicity conservation of $e+n\gamma_0$
particles before and $e+\gamma$ after the interaction \cite{GalSik92}.

Using (\ref{nely}) one can obtain an expression for the degree of the
longitudinal polarization of the final electron in the case when the final
photon polarization is not detected:
\beq
\lambda_e^{f} = \sum_{n=1}^{\infty}\;(\lambda_e\;F_{1n}+\lambda_e\;F_{5n}+\lambda\;
F_{2n})\;/\;\sum_{n=1}^{\infty} \;(\; F_{1n} + \lambda\lambda_{e}\;F_{2n}\;)\; .
\label{laef}
\eeq
Summing (\ref{nely}) over polarizations of the final electrons, one
obtains the differential cross section taking into account the polarizations
of three particles: \cite{Rek83,GalSik92,Tsai,KGL,GalSik98}:
\beq
{d \sigma_c \over d y}(\lambda, \lambda_e, \lambda') =
{ \pi \alpha^2 \over x m^2 \xi^2 } \sum_{n = 1} ^{\infty}
(\;F_{1n} +\lambda \lambda_{e} F_{2n} + \lambda \lambda' F_{3n} +
 \lambda_e \lambda' F_{4n}\;)\; .
\label{nely3}
\eeq
This formula allows one to find the degree of the circular
polarization of the Compton photon $\lambda_{\gamma}^{f}$ for the case
when the final electron polarization is not detected:
\beq
\lambda_{\gamma}^{f} = \sum_{n=1}^{\infty} \; ( \lambda \; F_{3n}
+ \lambda_{e} \; F_{4n} \; ) \; / \; \sum_{n=1}^{\infty} \; ( \;  F_{1n}
+  \lambda \lambda_{e} \; F_{2n} \; )  \; .
\label{lagf}
\eeq
After summation of (\ref{nely3}) over polarizations of the final
photons one gets the differential cross section of BCS in which
polarizations of the initial particles are taken into account:
\beq
{d \sigma_c \over d y}(\lambda, \lambda_e) =
{2 \pi \alpha^2 \over x m^2 \xi^2 }
 \sum_{n = 1} ^{\infty} \;(\;F_{1n}
+\lambda \lambda_{e} F_{2n} \;)\; .
\label{nely2}
\eeq
The photon energy spectra $f(x,y)$ are defined through the differential
cross section ${d \sigma_c}(\lambda, \lambda_e)/dy$
(\ref{nely2}) (for simplicity, we suppress below in all expressions the
indexes $\lambda, \lambda_e$):
\beq
f(x,y) \equiv {1 \over \sigma_c}  \; { d \sigma_c \over dy}\, ,\; \;
\sigma_c={2 \pi \alpha^2 \over x m^2 \xi^2 }\sum_{n=1}^{n_{max}}\int_0^{y_n}
 (\;F_{1n} +\lambda \lambda_{e} F_{2n} \;) \,dy\; ,
\label{difspec}
\eeq
where $\sigma_c$ is the total cross section,  $n_{max}\equiv n_m$ can
be found in the condition that the series (\ref{nely2}) converges.
It defines the upper edge of the spectrum $y_{max} \equiv y_m=n_m x
/(n_m x+1+\xi^2)$ in the nonlinear BCS (see (\ref{y_n})).

In the case of hard $\gamma$ quanta  colliding just after Compton
conversion, the distribution of the spectral luminosity
$\GG$ collisions $L_{\GG}$ over the invariant mass of colliding
photons $W_{\GG}=\sqrt{4\omega_1 \omega_2},\, z=W_{\GG}/2\varepsilon$
is expressed through the energy spectra of the photons
(\ref{difspec}):
\beq
{1\over {L}_{\GG}} {d {L}_{\GG} \over dz}=2z\; \int_{-\eta_m}
^{+\eta_m}f(x,ze^{+\eta})\;f(x,ze^{-\eta})\;d\eta \;,\; \;
\eta \equiv \ln \sqrt{y_1/y_2}\;,
\label{lumz}
\eeq
where $\eta$ is the rapidity of the $\GG$ system, $y_i=\omega_i/\varepsilon$
are parts of energies which are taken by photons moving opposite directions
1 and 2. Here $z$ varies between 0 and $z_m=W_m/2 \varepsilon=y_m$ with
$W_m$ being the maximum value of the invariant mass of the colliding photons
$W_{\GG}$, $W_m=2\omega_m$, and $\eta$ lies in the region $|\eta| \leq
\ln(y_m/z)$.

In the case of nonlinear BCS the photon scattering angle is unique
function of the photon energy analogous in its form to that for linear
Compton scattering: \beq \vartheta_{\gamma}^{n}(y)=\vartheta_0^n
\sqrt{{y_n \over y}-1}\;, \; \vartheta_0^n={m_{\ast} \over
  \varepsilon} \,\sqrt{1+u_n}\,,\, m_{\ast} =m\sqrt{1+\xi^2}\,.  \eeq
Below we analyze the influence of the nonlinear effect in BCS on the
energy spectra of photons $f(x,y)$ (\ref{difspec}), the helicities of
the final photon $\lambda_{\gamma}^{f}$ (\ref{lagf}) and electron
$\lambda_e^f$ (\ref{laef}) as well as on the luminosity spectra of
$\GG$ collisions (\ref{lumz}). Numerical calculations of these
quantities will be carried out at the following values of the
intensity parameter $\xi^2=0.0, 0.3, 1.0, 3.0$ and for $x=4.8, 20,
50$.  We will consider the three usual sets of the polarization
stations of the colliding particles: (i) the initial electron is
unpolarized ($\lambda_e =0$); ii) initial particles are completely
polarized ($\lambda_e=1, \lambda=\pm1$) and their spins are parallel
($\lambda \lambda_e=-1$); iii) the same for antiparallel spins
($\lambda \lambda_e=1$) (\ref{polnel})). In addition we will also
study the case when the electron is not completely polarized $\lambda_e=0.8$
(\ref{polreal}): \beqn 1) \to \lambda_{e} = 0 , \; \lambda &=& 1 ; \;
2) \to \lambda_{e} = 1 , \; \lambda = - 1 ; \; 3) \to \lambda_{e} = 1
\; , \; \lambda = 1 \, ,
\label{polnel}\\
 2r) & \to& \lambda_{e} = 0.8 ,
 \; \lambda = - 1 ; \; 3r) \to \lambda_{e} = 0.8 \; , \; \lambda = 1 \, ,
\label{polreal}
\eeqn
As a rule, BCS is considered only in the region $2.5<x<4.8$. At $x>4.8$
in the conversion region a process of $\EPEM$ pairs production becomes
possible at collisions of Compton photons with photons of the same
laser wave \cite{GKST83,TEL90}. The reaction threshold at $\xi^2=0$ is equal
to $\omega_m\omega_0 > m^2 $, i.e. $x = 2(1+\sqrt{2}) \approx 4.828$. Above
this threshold ($x \sim 8-20$) the cross section of the pair production
is 1.5--5 times larger than that of BCS. This leads to knocking out the
Compton photons and lowering the conversion coefficient \cite{TEL90,TEL95}.
Nevertheless, we will also consider the region of higher $x$ being of
interest for the experiments in which the maximum monochromatization of
$\GG$ collisions is required.

\section{Energy spectra of photons}

Results of numerical calculations of energy spectra of photons are shown
in Fig.1. It is seen in Fig.1 that the energy spectra of photons depend
strongly on the values of $\lambda \lambda_e$, $x$ and $\xi^2$. In the linear
case at $x=4.8$ when spins of colliding electrons and photons of the laser
wave are parallel ($\lambda \lambda_e=-1$), the number of high energy
photons is almost two times larger than that for the 
unpolarized case (see Fig.1a).
At  $x=50$ the yield of hard photons increases by a factor of three
(see Fig.1i). So, simultaneously with increasing $x$ (i.e. with increasing
$\omega_0$ or $\varepsilon$) there exits an effective ``pumping'' soft
photons into hard ones. This leads to noticeable improving the
monochromaticity of $\GG$ collisions. In contrast, if spins  of electrons
and laser photons are antiparallel ($\lambda \lambda_e=+1$) the number
of hard photons decreases ({\it cf.} curves 1 and 3 in Fig.1(a,e,i)).
Accordingly, the monochromaticity of collisions goes down.

Nonlinear effects in BCS ($\xi^2\neq 0$) lead to significant changes in
energy spectra in comparison with those in the linear BCS ($\xi^2= 0$).
First, simultaneous absorption of a few photons from a laser wave induces
widening spectra of hard $\gamma$ quanta and a generation of additional
peaks corresponding to radiation of higher harmonics. This widening at
the same parameter $x$ is more pronounced  the higher the intensity
of the wave ({\it cf.} Fig.1(a-d), (e-h), (i-l)).  A consequence of
the widening is the decrease of the height of the peak for the first
harmonics in comparison with that in the linear case. This is
clearly seen if one compares Figs.1(a-d) and so on.

Second, due to increase of the electron effective mass (\ref{ksi}) the
scattered photon have lower energies, i.e. the first harmonic is
shifted to lower values of $y$ (see Fig.1(a-d) and so on). 
With increasing the parameter $x$ the relative shift of the first
harmonic decreases \cite{GalSik92,TEL95}.

At a relatively small intensity of the laser wave ($\xi^2 \sim 0.3$)
the main contribution give  photons of the
first harmonic and the probability for generation of higher harmonics
is small (see Fig.1(b,f,j)). At $\xi \sim 1$ the widening the spectra
because of nonlinear effects is accompanied by an increase of the yield
of photons with energies higher than the maximum energy of the first harmonic
(see. Fig.1(c,g,k)). At last, at high intensities ($\xi^2 \ge 1$) the
nonlinear processes of scattering on many photon  can become
comparable with one photon scattering (see Fig.1d) and even  predominant
(at $\xi^2 \gg 1$) \cite{Ritus79,GalSik92}.

\section{Polarization of final photons}

Results of numerical calculations for the energy dependence of the degree
of the circular polarization of Compton photons (\ref{lagf}) are given
in Fig.2. We begin our consideration with the case of $\xi^2=0$ and
$\lambda\lambda_e=1$ (the case of bad monochromaticity). It is
seen in Fig.2(e,i), that at $x=20, 50$ the helicity of the Compton
photon is practically constant and equal to one in all energy range
$0<y<y_m$. As a result, the contribution of the state of the $\GG$
system with the spin 2 in the distribution of the total spectral
luminosity of $\GG$ collisions will be almost equal to zero. Note
that the increase of the intensity parameter $\xi^2$ up to 1 at
$x=50$ does not lead to noticeable changes in such a character of the
behavior of $\lambda_{\gamma}^f$. At the same time, by going to the
realistic electron polarizations ($\lambda_e=0.8$) the behavior of the
helicity of Compton photons (see Fig.2(i-l)) becomes  more
pronounced.

When the monochromaticity is best ($\lambda\lambda_e=-1$), both at
$\xi^2=0$ and $\xi^2 \neq 0$ there exists a rather wide energy interval near
$y \sim y_m$ where the scattered photons have a high degree of the
polarizations (almost 100 \%). If the initial electron polarization is
less than 100 \% and $\lambda=1$, the energy range of the high helicity
becomes markedly narrower. High energy photons can have a high
degree of circular polarization near $y \sim y_m$ at $\lambda=1$ even
if the initial electron is unpolarized ($\lambda_e=0$).

Nonlinear effects induce additional peaks in the energy dependence of
the helicity of the Compton photons. The height of the first peak which
corresponds to the radiation of the first harmonic decreases in magnitude
in comparison with the case of liner BCS and moves to lower energies.
It is worth mentioning that at small $y$ the helicity $\lambda_{\gamma}^f$
of the scattered photons is practically independent of the electron
polarization $\lambda_e$.

\section{Polarization of the final electrons}

Our results for the dependence of the degree of longitudinal polarization
of the scattered electron (\ref{laef}) on its energy $\varepsilon'/
\varepsilon=1-y$ are shown in Fig.3.
It is seen in Fig.3a for $x=4.8$ and $\xi^2=0$ that an unpolarized electron
being scattered from a totally polarized laser beam can acquire high
a degree of  longitudinal polarization (94 \%) in the region of
minimal values of the energy $\varepsilon'$. This is in agreement
with the result of a paper \cite{Kot}. In the case of linear BCS at
$x=20, 50$ the degree of the polarization of the final electron can
reach almost 100 \% (see Fig.3(e,i)).

Because of nonlinear effects, instead of one peak seen in
Fig.3(a,e,i) there emerge a few (for example, three at
$\xi^2=0.3$). The height of the first peak corresponding to the
first harmonic decreases in magnitude in comparison
with the case of linear BCS and moves to higher energies. At higher
intensities of the wave when $\xi^2=1.0, 3.0$ (see Fig.
3(c,d,g,h,k,l)) one has rather complicate a behavior due to the
absorption of many photons. 

\section{Distribution of total spectral luminosity of $\GG$ collisions }

  The total spectral luminosity of $\GG$ collisions
(\ref{lumz}) are shown in Fig.4. The luminosity spectra in Fig.4(a,e,i)
and the energy spectra in Fig.1(a,e,i) are consistent with each other.
It should be noted that in the approximation when the intensity of
the laser wave can be neglected, the distributions of the spectral
luminosity over the invariant mass have a sharp edge at $z=z_m$. This
can be very important in the search of the Higgs boson \cite{TELee}.

It is seen in Fig.4 that the behavior of the spectral luminosities
is analogous to that of photon energy spectra. Together with increasing
the wave intensity we observe the raising of the luminosity  in
the region of low and intermediate invariant masses of colliding photons,
and the decreasing of the height of the peaks. There emerges a long ``tail''
in the range of high invariant masses. Widening the luminosity distribution
in comparison with linear BCS leads to the appearance of small ``triangles''
located on the right side of the edge of the spectrum $\xi^2=0$. The slope
of the curves for the luminosity with respect to the abscissa axis
essentially changes. As a result, the sharp edge of the spectral
luminosity, which is inherent in linear BCS, disappears which leads
to noticeable decrease of the monochromaticity of $\GG$ collisions. All
unfavorable moments due to nonlinear effects are less important when
the parameter $x$ increases. This is clearly seen in Fig.4 for
to $x=4.8$,$x=20$ and $x=50$.

\section{The distribution of spectral luminosities for a $\GG$ system with
spins 0 and 2.}

 The spectral luminosity of $\GG$ collisions in the
case of spins 0 and 2 of the $\GG$ system, $dL_0/dz$ and $dL_2/dz$,
are shown in Fig.5 (for $\lambda\lambda_e=0,\pm1$) and Fig.6 (for
$\lambda\lambda_e=\pm0.8$).  Our calculations for linear BCS at
$\lambda\lambda_e=-1$ show that near the peak, i.e. for $0.9z_m \le z
\le z_m$, the main contribution to the total luminosity stems from the
state with the spin 0 (see Fig.5(b,e,h)). It was noted in papers
\cite{TEL95,Borden}, that the smallness the ratio $dL_2/dL_0$ at
$x=4.8$ in the peak region can be very important in search of the
Higgs boson. If the electron polarization is less 100 \%
($\lambda\lambda_e =-0.8$) then the contribution of the spin 2 states
grows.

For $\lambda \lambda_e=1$ and $\xi^2=0$ the ratio $dL_2/dL_0$ is also
small but in the full region $0 \le z \le z_m$ for all $x$ (see Fig.5(c,f,i)).
One can see also that with increasing $x$ the contribution of
the spin 2 state in the total luminosity is decreasing. It can be explained
by the fact that at high $x$ and $\lambda\lambda_e=1$ the scattered photon
is almost totally polarized (see Fig.2(e,i) and the discussion above).
At a collision of two such photons moving in reverse directions,
only the total spin 0 system is possible. Since in the case under
consideration nonlinear effects have small influence on the polarization
of final photons the suppression above will take place at high
intensities too (see Fig.5(f,i)). Contribution of the spin 2 state
significantly increase when the electron polarization $\lambda_e\ < 1$
(see Fig.6(b,d,f)). 

\section{Conclusion}

The process of backward Compton scattering is the best method for
obtaining high energy photons at future photon colliders.  The the
required density of laser photons in the conversion region is so high
that nonlinear effects in the Compton scattering take place.  The lead
to some increase of the maximum energy of scattered photons, however,
when the intensity of the laser wave grows the monochromaticity of
$\GG$ collisions becomes worse and the spectral luminosity is shifted
to the region of low and intermediate invariant masses and less
photons remain in the high energy peak. This is due the fact that only
photons of the first harmonic can be emitted along the direction of
the initial electron beam and produce photons near kinematical limit.
Such a behavior is not permitted for photons of higher order harmonics
because of the helicity conservation for the particle system
$e+n\gamma_0$ before interaction and $e+\gamma$ after interaction
\cite{GalSik92}. This leads to widening of the angular distribution of
the higher order harmonics and in turn decreases the monochromaticity
of $\GG$ collisions.

In this paper formulae for energy distributions of final photons end
electrons and their polarizations are given in the case of nonlinear
Compton scattering. Numerous figures for energy spectra and luminosity
distributions at different values of initial beam polarizations and parameters
$x$ and $\xi^2$ are presented, which allows to see main tendencies.

\section{Appendix}

In the present paper the numerical calculations of energy spectra,
polarizations of final electrons and photons and luminosities are
carried out making the use of precise formulae for the corresponding
differential cross sections. However, for small intensities when
$\xi^2\ll 1$ one can restrict oneself to the contribution of two first
harmonics to the differential cross sections (\ref{nely}), (\ref{nely3}),
and (\ref{nely2}) and obtain for them approximate expressions. To this end,
it is quite enough to expand the Bessel functions in (\ref{nely}) in terms
of the parameter $\Delta=\xi^2/(1+ \xi^2)$ (rather than in terms of $\xi^2$
as is done in \cite{Ritus}) and for $u_n\, (y_n)$ (\ref{y_n}) to use
exact formulae. Only in this way can one  reveal the above pointed out
kinematical features of the functions (\ref{funy}) $F_{in}\,
(i=1,\ldots5)$. As a result, we have (see \cite{GalSik92}):
\beqn
{F_{11} \over \xi^2}&=&1-y+{1 \over 1-y}-4 \alpha_1 -4 \alpha_1 \Delta\;
\Bigg(-y +{1 \over 1-y} - \alpha_1\Bigg) \nonumber, \\
{F_{21} \over \xi^2}&=&\Bigg(-1+y+{1 \over 1-y}\Bigg)\Bigg(1- 2{y \over y_1}
{(1-y_1) \over (1-y)}\Bigg) \big(1-2\alpha_1 \,\Delta\big)\; ,\nonumber\\
{F_{31} \over \xi^2}&=&\Bigg(1-y+{1 \over 1-y}\Bigg)\Bigg(1- 2{y \over y_1}
{(1-y_1) \over (1-y)}\Bigg) \big(1-2\alpha_1 \,\Delta \big)\; ,
\label{1harm} \\
{F_{41} \over \xi^2}&=&-1+y+{1 \over 1-y}-4 \alpha_1\,y -4 \alpha_1 \Delta\,y
\Bigg(- \alpha_1+{1 \over 1-y} \Bigg) \nonumber,\\
{F_{51} \over \xi^2}&=&4 \alpha_1(1-\Delta(1+ \alpha_1))\,\Bigg(1+y-
{1 \over 1-y}\Bigg)\, ,\nonumber
\eeqn
for the first harmonic;
\beqn
{F_{12} \over \xi^2}&=&4 \alpha_2 \Delta\,
\Bigg(1-y +{1 \over 1-y} -4 \alpha_2\Bigg) \nonumber, \\
{F_{22} \over \xi^2}&=&4 \alpha_2 \Delta\, \Bigg(-1+y+{1 \over 1-y}\Bigg)
\Bigg(1- 2{y \over y_2} {(1-y_2) \over (1-y)}\Bigg) ,\nonumber\\
{F_{32} \over \xi^2}&=&4 \alpha_2 \Delta\, \Bigg(1-y+{1 \over 1-y}\Bigg)
\Bigg(1- 2{y \over y_2} {(1-y_2) \over (1-y)}\Bigg) ,
\label{2harm} \\
{F_{42} \over \xi^2}&=&4 \alpha_2 \Delta\,
\Bigg(-1+y +{1 \over 1-y} -4 \alpha_2y\Bigg) \nonumber, \\
{F_{52} \over \xi^2}&=&16 \alpha_2^2\Delta \,\Bigg(1+y-{1 \over 1-y}\Bigg)
\, ,\nonumber
\eeqn
for the second harmonic.

The total cross section of the process (\ref{nelcom})
$\sigma_c (\lambda, \lambda_e)$ (\ref{difspec}) has the following form
\cite{GalSik92}:
\beq
\sigma_c (\lambda, \lambda_e)={2 \pi \alpha^2 \over x m^2 }
\;(\;f_1+\lambda \lambda_{e}\,f_2\,)\, ,
\,f_1=f_{11}+f_{12}\, , f_2=f_{21}+f_{22}\, ,
\label{nel2g}
\eeq
where
\beqn
f_{11}&=&\Bigg(1-{4 \over u_1}-{8 \over u_1^2}\Bigg) \ln(1+u_1)+{1 \over 2}
+{8 \over u_1} -{1 \over 2 (1+u_1)^2}
-\Delta \Bigg(2+{44 \over 3 u_1} -\nonumber\\
&-&{16 \over u_1^2}-{16 \over u_1^3}
-{2 \over 1+u_1} -{8 \over u_1}\Bigg(1+{1 \over u_1}-{3 \over u_1^2}
-{2 \over u_1^3}\Bigg) \ln(1+u_1)\Bigg)\,, \nonumber\\
f_{21}&=&\Bigg(1+{2 \over u_1} \Bigg)\ln(1+u_1) -{5 \over 2} +{1 \over 1+u_1}
-{1 \over 2 (1+u_1)^2}-\nonumber \\
&-&\Delta\Bigg({1 \over 3} +{4 \over u_1}-{8 \over u_1^2}
-{1 \over 1+u_1} -{2 \over u_1}\Bigg(1-{4 \over u_1^2}\Bigg)
\ln(1+u_1)\Bigg)\,, \label{fitot}\\
f_{12}&=&4\Delta \Bigg({1 \over 2} +{2 \over 3 u_2}-{16 \over u_2^2}
-{16 \over u_2^3} -{1 \over 2 (1+u_2)}-\nonumber\\
&-&{1 \over u_2}\Bigg(1-{6 \over u_2}-{24 \over u_2^2} -{16 \over u_2^3}
\Bigg) \ln(1+u_2) \Bigg)\,,\nonumber\\
f_{22}&=&4\Delta\Bigg({1 \over 6}+{2 \over u_2}-{4 \over u_2^2}-{1 \over
2(1+u_2)}-{1 \over u_2}\Bigg(1-{4 \over u_2^2}\Bigg) \ln(1+u_2) \Bigg)\,.
\nonumber
\eeqn
Note that computations of the energy spectra of photons for $x=4.8$
and $\xi^2=0.3$ done with the use either of exact formulae
(\ref{difspec}) or approximated ones (\ref{1harm}), (\ref{2harm}) and
(\ref{nel2g}) lead to practically the same results. For example, the
difference in the magnitude of peaks corresponding to first harmonics
is only about 0.5 \%.

\newpage
\centerline{Figure Caption}
Fig.1
The energy spectra of Compton photons (\ref{difspec})
for various values of $x$, $\xi^2$ and $\lambda \lambda_e$. In
Fig.(a-d), (e-h), and (i-l) parameter $x=4.8, 20$, and $50$,
respectively. In Fig.(a,e,i), (b,f,j), (c,g,k), and (d,h,l)
the intensity parameter $\xi^2=0.0, 0.3, 1.0$, and $3.0$,
respectively.  Lines labelled by 1, 2, 3 correspond to the choice of
polarization stations of initial particles $\lambda_e=0, \lambda=1$;
$\lambda_e=1, \lambda=-1$; $\lambda_e=1, \lambda=1$.

Fig.2 The energy dependence of the helicity of Compton photons
(\ref{lagf}) for various values of $x$, $\xi^2$ and $\lambda
\lambda_e$. In Fig.(a-d), (e-h), and (i-l) parameter $x=4.8, 20$, and
$50$, respectively. In Fig.(a,e,i), (b,f,j), (c,g,k), and (d,h,l) the
intensity parameter $\xi^2=0.0, 0.3, 1.0$, and $3.0$, respectively.
The full, long-dash-dotted, and dashed curves correspond to the set of
polarization states $\lambda_e=0, \lambda=1$; $\lambda_e=1,
\lambda=-1$; $\lambda_e=1, \lambda=1$, respectively.
Short-dash-dotted and dotted curves correspond to choice
$\lambda_e=0.8, \lambda=-1$ and $\lambda_e=0.8, \lambda=+1$,
respectively.

Fig.3 The degree of the longitudinal polarization of final electron
(\ref{laef}) as a function of its energy $\varepsilon'/
\varepsilon=1-y$ for various values $x$, $\xi^2$, and $\lambda
\lambda_e $. In Fig.(a-d), (e-h), and (i-l) the parameter $x=4.8, 20$
and $50$, respectively. In Fig.(a,e,i), (b,f,j), (c,g,k), and (d,h,l)
the intensity parameter $\xi^2=0.0, 0.3, 1.0$, and $3.0$,
respectively. In Fig.(a-l) the solid, dotted, and dashed curves
correspond to $\lambda_e=0, \lambda=1$; $\lambda_e=1, \lambda=-1$;
$\lambda_e=1, \lambda=1$, respectively.

Fig.4 The distribution of total spectral luminosity of $\GG$
collisions (\ref{lumz}) over the invariant mass of photons
$z=W_{\GG}/2\varepsilon$.  In Fig.(a-d), (e-h), and (i-l) parameter
$x=4.8$, 20, and 50, respectively. In Fig.(a,e,i), the intensity
parameter $\xi^2=0.0$, and the curves 1, 2, and 3 stand for the choice
of the polarizations $\lambda_e=0, \lambda=1$; $\lambda_e=1,
\lambda=-1$; $\lambda_e=1, \lambda=1$. The curves in Fig.(b,f,j),
(c,g,k), and (d,h,l) correspond to $\lambda\lambda_e=0$, $\lambda
\lambda_e=-1$, and $\lambda\lambda_e=1$, respectively. In all these
figures except for (a,e,i) the solid, dotted, dashed, and dash-dotted
curves stand for the intensity parameter $\xi^2$ equal to 0.0, 0.3,
1.0, and 3.0, respectively.

Fig.5
The distribution of spectral luminosities $dL_0/dz$ and $dL_2/dz$ of the
$\GG$ system over the invariant mass of photons $z=W_{\GG}/2\varepsilon$.
Corresponding curves are labelled by numbers 0 and 2 (for spin 0 and 2).
The left, central, and right panels correspond to $x=4.8$, 20, and 50,
respectively. Fig.(a,d,g), (b,e,h), and (c,f,i) were built for
$\lambda\lambda_e=0$, $\lambda\lambda_e=-1$, and $\lambda\lambda_e=1$,
respectively. The solid, dotted, dashed, and dash-dotted curves
correspond to the intensity parameter $\xi^2=0.0, 0.3, 1.0$, and
$3.0$, respectively.

Fig.6
The distribution of spectral luminosities $dL_0/dz$ and $dL_2/dz$ of the
$\GG$ system over $z=W_{\GG}/2\varepsilon$. Corresponding curves are labelled
by numbers 0 and 2. The left, central, and right panels correspond to
$x=4.8$, 20 and 50, respectively. Fig.(a,c,e) and (b,d,f) stand for
$\lambda \lambda_e=-0.8$ and $\lambda \lambda_e=+0.8$, respectively.
The solid, dotted, dashed, and dash-dotted curves correspond to the
intensity parameter $\xi^2=0.0, 0.3, 1.0$, and $3.0$, respectively.

\newpage
\begin{figure}[hbt] \label{spectr}
\centering
\vspace*{-0.5cm}
\epsfig{file=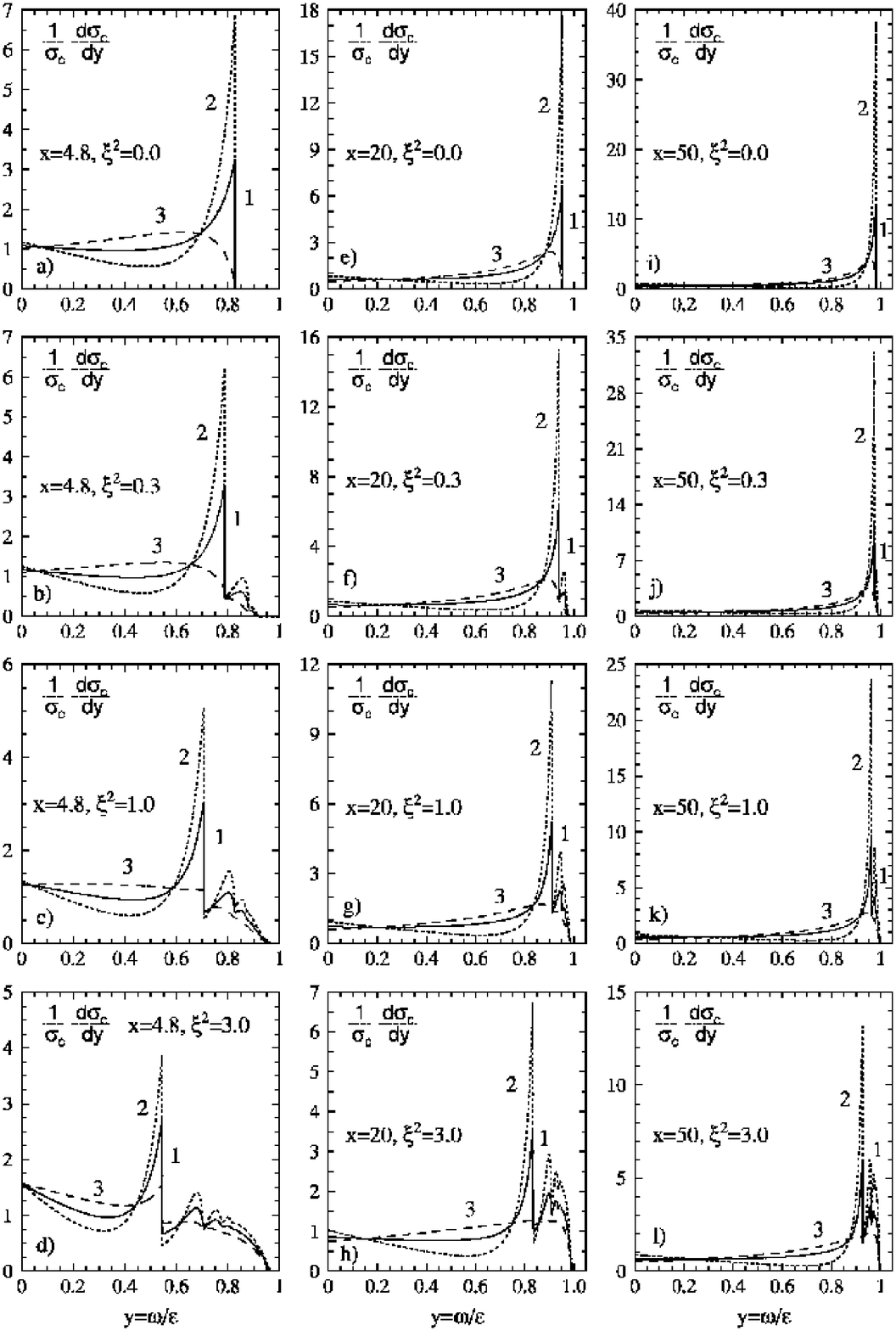,width=1.\textwidth}
\vspace{0cm}
\caption{
The energy spectra of Compton photons (\ref{difspec})
for various values of $x$, $\xi^2$ and $\lambda \lambda_e$. In
Fig.(a-d), (e-h), and (i-l) parameter $x=4.8, 20$, and $50$,
respectively. In Fig.(a,e,i), (b,f,j), (c,g,k), and (d,h,l)
the intensity parameter $\xi^2=0.0, 0.3, 1.0$, and $3.0$,
respectively.  Lines labelled by 1, 2, 3 correspond to the choice of
polarization stations of initial particles $\lambda_e=0, \lambda=1$;
$\lambda_e=1, \lambda=-1$; $\lambda_e=1, \lambda=1$.
}
\end{figure}

\begin{figure}[thb]
\vspace{-0.8cm}
\centering
\vspace*{0.2cm}
\epsfig{file=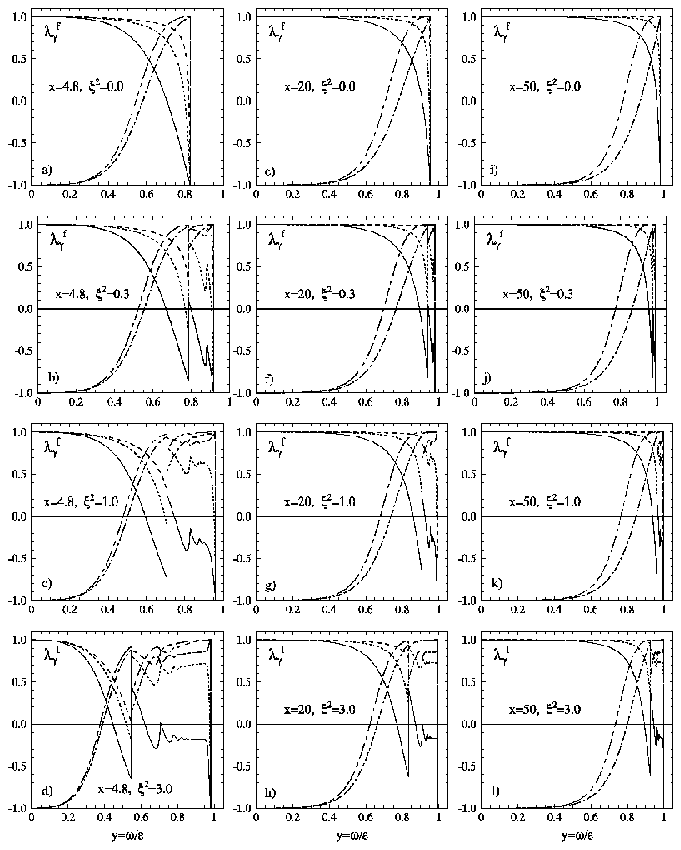,width=1.1\textwidth}
\caption{
 The energy dependence of the helicity of Compton photons
(\ref{lagf}) for various values of $x$, $\xi^2$ and $\lambda
\lambda_e$. In Fig.(a-d), (e-h), and (i-l) parameter $x=4.8, 20$, and
$50$, respectively. In Fig.(a,e,i), (b,f,j), (c,g,k), and (d,h,l) the
intensity parameter $\xi^2=0.0, 0.3, 1.0$, and $3.0$, respectively.
The full, long-dash-dotted, and dashed curves correspond to the set of
polarization states $\lambda_e=0, \lambda=1$; $\lambda_e=1,
\lambda=-1$; $\lambda_e=1, \lambda=1$, respectively.
Short-dash-dotted and dotted curves correspond to choice
$\lambda_e=0.8, \lambda=-1$ and $\lambda_e=0.8, \lambda=+1$,
respectively.
}
\end{figure}

\begin{figure}[hbt]
\centering
\vspace*{0.2cm}
\epsfig{file=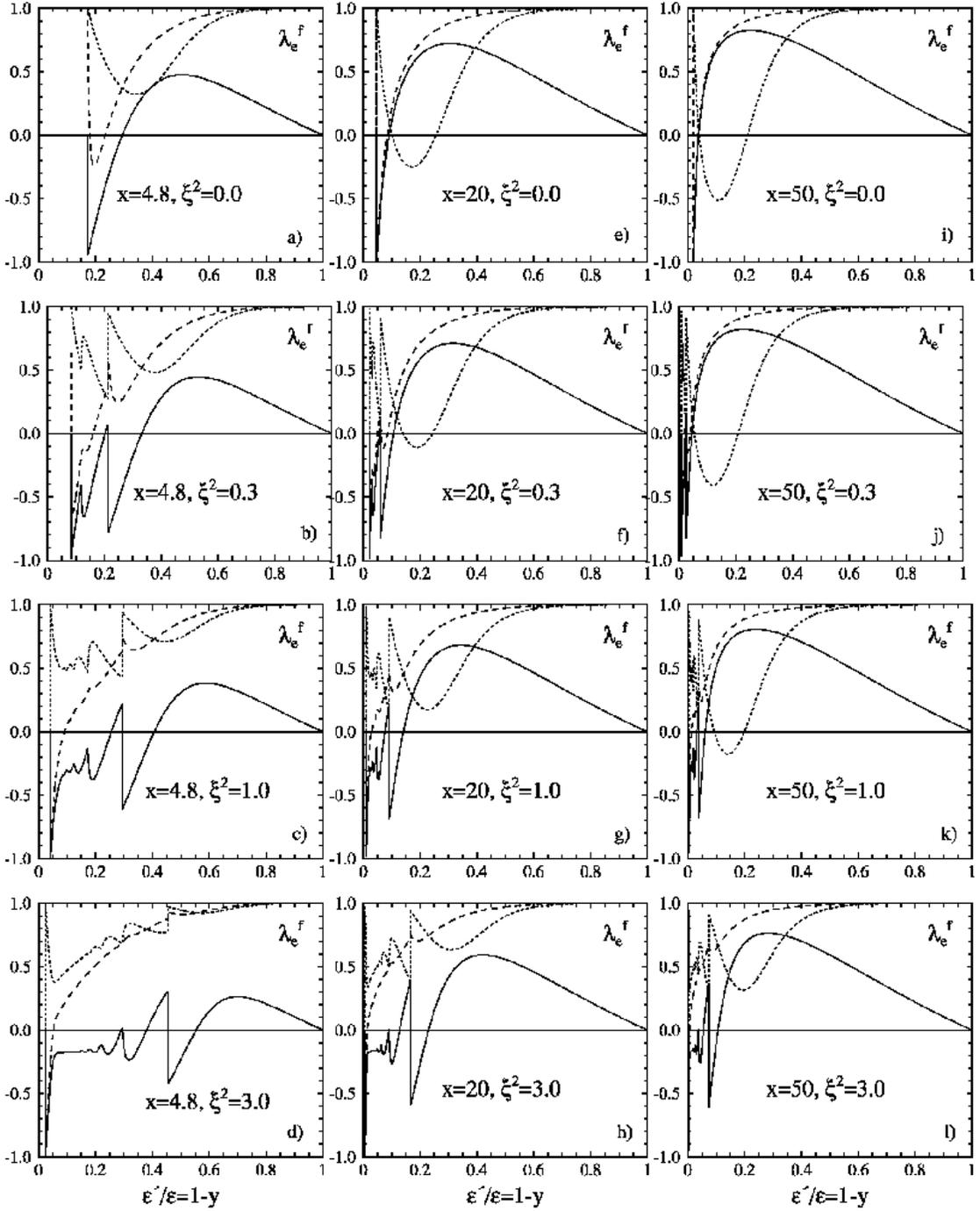,width=1.1\textwidth}
\vspace{-0.3cm}
\caption{
The degree of the longitudinal polarization of final electron
(\ref{laef}) as a function of its energy $\varepsilon'/
\varepsilon=1-y$ for various values $x$, $\xi^2$, and $\lambda
\lambda_e $. In Fig.(a-d), (e-h), and (i-l) the parameter $x=4.8, 20$
and $50$, respectively. In Fig.(a,e,i), (b,f,j), (c,g,k), and (d,h,l)
the intensity parameter $\xi^2=0.0, 0.3, 1.0$, and $3.0$,
respectively. In Fig.(a-l) the solid, dotted, and dashed curves
correspond to $\lambda_e=0, \lambda=1$; $\lambda_e=1, \lambda=-1$;
$\lambda_e=1, \lambda=1$, respectively.
}
\end{figure}

\begin{figure}[ht]
\centering
\vspace*{0.2cm}
\epsfig{file=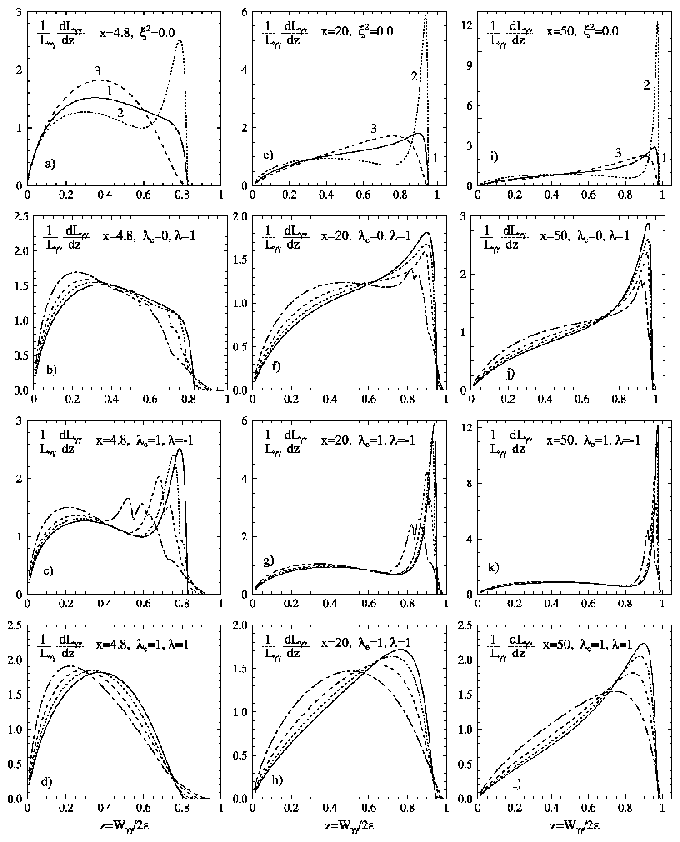,width=1.07\textwidth}
\caption{
The distribution of total spectral luminosity of $\GG$
collisions (\ref{lumz}) over the invariant mass of photons
$z=W_{\GG}/2\varepsilon$.  In Fig.(a-d), (e-h), and (i-l) parameter
$x=4.8$, 20, and 50, respectively. In Fig.(a,e,i), the intensity
parameter $\xi^2=0.0$, and the curves 1, 2, and 3 stand for the choice
of the polarizations $\lambda_e=0, \lambda=1$; $\lambda_e=1,
\lambda=-1$; $\lambda_e=1, \lambda=1$. The curves in Fig.(b,f,j),
(c,g,k), and (d,h,l) correspond to $\lambda\lambda_e=0$, $\lambda
\lambda_e=-1$, and $\lambda\lambda_e=1$, respectively. In all these
figures except for (a,e,i) the solid, dotted, dashed, and dash-dotted
curves stand for the intensity parameter $\xi^2$ equal to 0.0, 0.3,
1.0, and 3.0, respectively.
}
\end{figure}

\begin{figure}[hbt]
\centering
\vspace*{0.2cm}
\epsfig{file=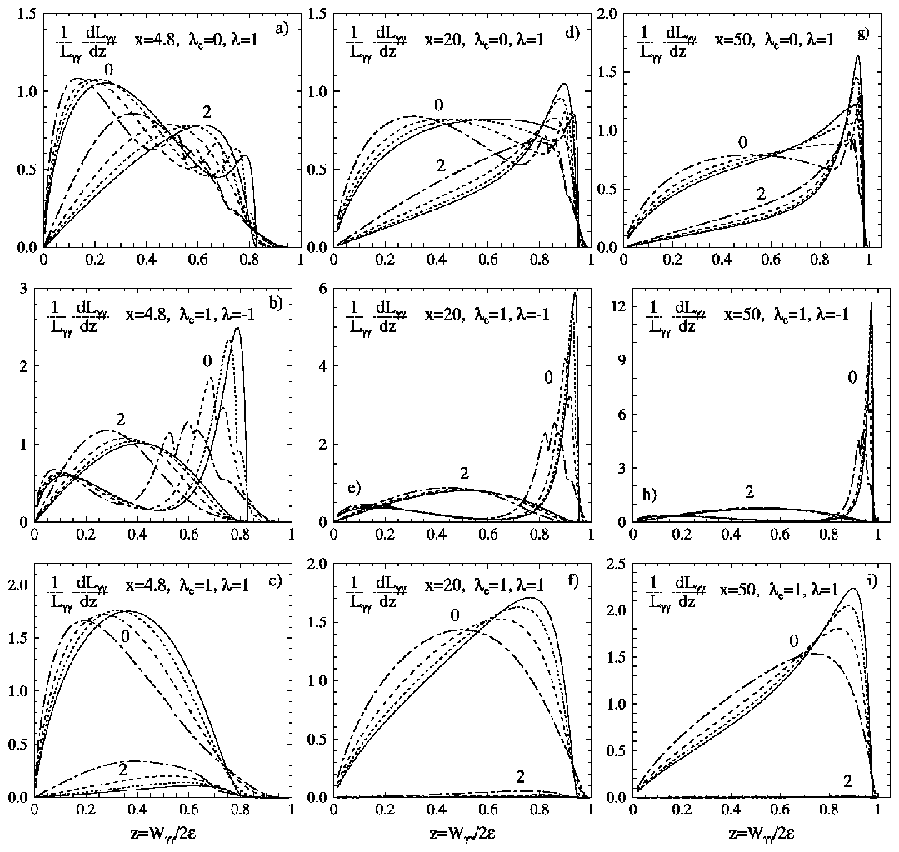,width=1.1\textwidth}
\vspace{-0.30cm}
\caption{
The distribution of spectral luminosities $dL_0/dz$ and $dL_2/dz$ of the
$\GG$ system over the invariant mass of photons $z=W_{\GG}/2\varepsilon$.
Corresponding curves are labelled by numbers 0 and 2 (for spin 0 and 2).
The left, central, and right panels correspond to $x=4.8$, 20, and 50,
respectively. Fig.(a,d,g), (b,e,h), and (c,f,i) were built for
$\lambda\lambda_e=0$, $\lambda\lambda_e=-1$, and $\lambda\lambda_e=1$,
respectively. The solid, dotted, dashed, and dash-dotted curves
correspond to the intensity parameter $\xi^2=0.0, 0.3, 1.0$, and
$3.0$, respectively.
}
\end{figure}

\begin{figure}[htb]
\centering
\vspace*{0.2cm}
\epsfig{file=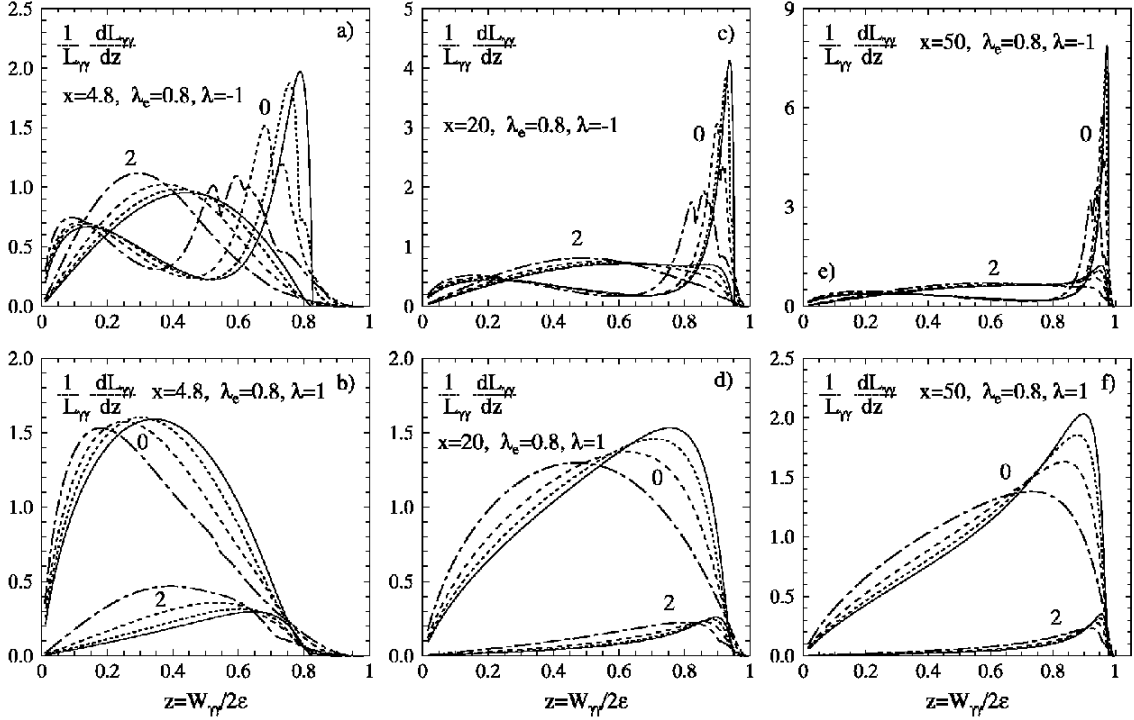,width=1.1\textwidth}
\vspace{-0.30cm}
\caption{
The distribution of spectral luminosities $dL_0/dz$ and $dL_2/dz$ of the
$\GG$ system over $z=W_{\GG}/2\varepsilon$. Corresponding curves are labelled
by numbers 0 and 2. The left, central, and right panels correspond to
$x=4.8$, 20 and 50, respectively. Fig.(a,c,e) and (b,d,f) stand for
$\lambda \lambda_e=-0.8$ and $\lambda \lambda_e=+0.8$, respectively.
The solid, dotted, dashed, and dash-dotted curves correspond to the
intensity parameter $\xi^2=0.0, 0.3, 1.0$, and $3.0$, respectively.
}
\end{figure}
\end{document}